\documentclass[aip,physrev,preprint]{revtex4-2} 

\usepackage{soul}
\usepackage{graphicx}
\graphicspath{{figures/}}
\usepackage{physics}
\usepackage{multirow}
\usepackage{xcolor}
\usepackage{caption}
\usepackage{subcaption}
\usepackage{bm}
\usepackage{amsfonts}
\usepackage{yfonts} 
\usepackage[cal=boondoxo,bb=stix]{mathalpha}  
\usepackage[title,titletoc]{appendix}

\usepackage{diagbox}
\usepackage{booktabs} 
\usepackage{threeparttable}
\usepackage{float}

\usepackage{xspace} 

\usepackage{siunitx}
\usepackage{hyperref} 
\hypersetup{colorlinks=true, citecolor=blue, urlcolor=blue, linkcolor=blue}
\usepackage{cleveref}
  	\crefname{figure}{Figure}{Figures}
  	\crefname{table}{Table}{Tables}
  	\crefname{equation}{Eq.}{Eqs.}
  	\crefname{section}{Section}{Sections}
  	\crefname{subsection}{Section}{Sections}
  	\crefname{subsubsection}{Section}{Sections}
  	\crefname{algorithm}{Algorithm}{Algorithms}
    \crefname{appendix}{Appendix}{Appendices}

\newcommand{\code}[1]{\texttt{#1}}
\newcommand\vartextvisiblespace[1][.5em]{%
  \makebox[#1]{%
    \kern.07em
    \vrule height.3ex
    \hrulefill
    \vrule height.3ex
    \kern.07em
  }
}

\usepackage{todonotes}

\newcommand{\ga}{\ensuremath{\mathbf{a}}}
\newcommand{\gb}{\ensuremath{\mathbf{b}}}

\newcommand{\gzeroA}{\ensuremath{\bm{0}_A}}
\newcommand{\gzeroB}{\ensuremath{\bm{0}_B}}
\newcommand{\gzeroC}{\ensuremath{\bm{0}_C}}

\begin{document}

\title{Comment on ``Efficient implementation of the superposition  of atomic potentials initial guess for electronic  structure calculations in Gaussian basis sets''} 

\author{Kshitijkumar A. Surjuse}
\author{Zhihao Deng}
\author{Andrey Asadchev}
\author{Edward F. Valeev}
\email{efv@vt.edu}
\affiliation{Department of Chemistry, Virginia Tech, Blacksburg, VA 24061}

\date{\today}
\newpage

\begin{abstract}
\end{abstract}

\maketitle 

In Ref. \citenum{VRG:lehtola:2020:JCP} Lehtola, Visscher, and Engel reported how the matrix representation of Superposition-of-Atomic-Potentials (SAP) potential\cite{VRG:lehtola:2019:JCTC} in Gaussian AO basis (``SAP matrix'') can be obtained efficiently by expressing the electronic contribution to SAP as the electrostatic potential of (contracted) spherically-symmetric Gaussians. This allows one to compute the SAP matrix using the conventional Gaussian AO integral machinery, nearly ubiquitous in molecular and increasingly solid-state computations. The authors noted this in the abstract: ``The guess obtained from the fitted potentials can be easily implemented in any Gaussian-basis quantum chemistry code in terms of two-electron integrals.'' The purpose of this Comment is to point out that the SAP matrix can be obtained as a trivial modification [\cref{eq:BoysReplacementSAP}] of the {\em Boys route} for {\em one-electron} nuclear attraction integrals, without the explicit use of two-electron integrals. Although our focus is on the Boys route, note that in the {\em Rys quadrature route} evaluation of the SAP matrix can also be accomplished by adjusting the quadrature roots/weights used in the one-electron nuclear attraction integral to refer to the SAP kernel [\cref{eq:v-sap-r3}]. 

The Boys route\cite{VRG:boys:1950:PRSMPES} expresses 1- and 2-particle Gaussian AO integrals involving non-separable Poisson kernel $|\mathbf{r}|^{-1}$ (such as the electron-nucleus and electron-electron integrals) as a linear combination of 1-dimensional integrals (Boys functions for the Coulomb integrals, and other similar integrals for non-Coulomb integrals\cite{VRG:ahlrichs:2006:PCCP}).
Among the many evaluation schemes that follow the Boys route, we will only mention the two major ones: the Obara-Saika\cite{VRG:obara:1986:JCP} and McMurchie-Davidson\cite{VRG:mcmurchie:1978:JCP}, each encompassing a large number of variants\cite{VRG:head-gordon:1988:JCP,VRG:ahlrichs:2004:PCCP,VRG:gill:1991:IJQC}.

We will closely follow the notation and definitions of the original Obara-Saika (OS) paper\cite{VRG:obara:1986:JCP}.
Contribution to the Gaussian-fitted SAP potential from the nucleus $C$ is expressed as a sum of the nuclear contributions, defined by the electrostatic potential of the point charge $Z_C$, and the electronic contribution expressed as the electrostatic potential of the generating ``density'':\cite{VRG:lehtola:2020:JCP}
\begin{align} \label{eq:v-sap-r2}
    V_C^{\text{SAP}}(\mathbf{r}) \equiv & V_C(\mathbf{r}) + \int \mathrm{d}\mathbf{r}' \, \frac{\theta_C(\mathbf{r}')}{|\mathbf{r} - \mathbf{r}'|}, \\
    V_C(\mathbf{r}) \equiv & -  \frac{Z_C}{r_C} 
\end{align}
where $r_C \equiv |{\bf r}-{\bf C}|$, and ${\bf C}$ is the position of the nucleus $C$.

The generating density,
\begin{align}
\label{eq:thetaC}
\theta_C(\mathbf{r}) \equiv & \sum_k c_k \tilde{s}_{\alpha_k}(r_C),
\end{align}
is a superposition of unit-charge (not unit-norm) spherical Gaussians:
\begin{align}
\tilde{s}_{\alpha}(r) \equiv &\left(\frac{\alpha}{\pi}\right)^{3/2} \exp(-\alpha r^2 ).
\end{align}
The coefficients and exponents of the unit-charge Gaussians, $\{c_k,\alpha_k\}$, are pre-tabulated for each atomic number $Z_C$.\cite{VRG:lehtola:2020:JCP}
Since the electrostatic potential of $\tilde{s}_{\alpha}$ is $\erf(\sqrt{\alpha} r)/r$,
$V_C^{\text{SAP}}$ can be rewritten as
\begin{align} \label{eq:v-sap-r3}
    V_C^{\text{SAP}}(\mathbf{r}) \equiv & - Z_C \frac{1 - \sum_k \tilde{c}_k \erf(\sqrt{\alpha_k} r_C) }{r_C},
\end{align}
using renormalized fitting coefficients
\begin{align}
    \tilde{c}_k \equiv \frac{c_k}{Z_C}.
\end{align}
Gaussian integrals with kernels similar to  \cref{eq:v-sap-r3} are well-known in electronic structure theory,\cite{VRG:gill:1996:CPL} e.g. in the context of range-separated density functional theory, where kernels with a single attenuation parameter are encountered:
\begin{align}
\label{eq:rsX}
V_\text{rsX}(r) = \frac{a - b \erf(\omega r)}{r},
\end{align}
whereas the SAP kernel contains many such parameters $\alpha_k$. In the context of the OS scheme for 2-electron integrals the kernel for \cref{eq:rsX} with $a=b=1$ was considered by Ahlrichs.\cite{VRG:ahlrichs:2006:PCCP} As a trivial consequence of Ahlrichs' result we can show that in the Boys route the formulas for the SAP integral 
\begin{align}
(\mathbf{a}|V_C^\text{SAP}|\mathbf{b}) \equiv \int \phi^*_\mathbf{a}(\mathbf{r})  \,V_C^\text{SAP}(\mathbf{r}) \, \phi_\mathbf{b}(\mathbf{r}) \, \mathrm{d}\mathbf{r}
\end{align}
over primitive Gaussian AOs
\begin{align}
    \phi_\ga ({\bf r}) \equiv (x - A_x)^{a_x} (y - A_y)^{a_y} (z - A_z)^{a_z} \exp(-\zeta_a |{\bf r} - {\bf A}|^2).
\end{align}
can be obtained from the formulas for the corresponding 1-electron integrals for the electrostatic potential of the nucleus $C$,
\begin{align}
    (\mathbf{a}|V_C|\mathbf{b}) \equiv \int \phi^*_\mathbf{a}(\mathbf{r})  \,  \frac{-Z_C}{\vert \mathbf{r} - \mathbf{C} \vert} \, \phi_\mathbf{b}(\mathbf{r}) \, \mathrm{d}\mathbf{r},
\end{align}
by the following replacement of the bare Boys function values by their SAP-dressed counterpart:
\begin{align}
\label{eq:BoysReplacementSAP}
\underbrace{F_m(T)\vphantom{\sum_k\left( \frac{\alpha_k}{\zeta + \alpha_k} \right)^{m+\frac{1}{2}}}}_{V_C} \quad \to \quad \underbrace{F_m(T) - \sum_k \tilde{c}_k \left( \frac{\alpha_k}{\zeta + \alpha_k} \right)^{m+\frac{1}{2}} F_m\left(T \frac{\alpha_k}{\zeta + \alpha_k} \right)}_{V^\text{SAP}_C},
\end{align}
where the usual definitions have been used:
\begin{align}
\zeta \equiv & \zeta_a + \zeta_b \\
\mathbf{P} \equiv & \frac{\zeta_a \mathbf{A} + \zeta_b \mathbf{B}}{\zeta} \\
T \equiv & \zeta \vert \mathbf{P} - \mathbf{C}\vert^2 \\
F_m(x) \equiv & \int_0^1 t^{2m} \exp(- x t^2) \, \mathrm{d}t.
\end{align}

It may seem counterintuitive to the readers familiar with the Obara-Saika scheme that it is possible to ``fuse'' the evaluation of 3-center 2-electron integrals with the nuclear attraction integrals, since their recurrence relations are seemingly different.
Consider the OS (vertical) recurrence relation for integrals of the electrostatic potential $V_C$ due to nucleus  $C$ (Eq. (A19) in Ref. \citenum{VRG:obara:1986:JCP}),
\begin{align}
\label{eq:vrr_ab_V}
    (\ga + \bm{1}_i | V_C | \gb )^{(m)} = & (\mathbf{PA})_i (\ga  | V_C | \gb)^{(m)} -
    (\mathbf{PC})_i (\ga |V_C | \gb)^{(m+1)} \nonumber \\
    & + \frac{a_i}{2\zeta} \Bigl( (\ga - \bm{1}_i |V_C | \gb)^{(m)} - (\ga - \bm{1}_i |V_C|\gb)^{(m+1)} \Bigr) \nonumber \\
    & + \frac{b_i}{2\zeta} \Bigl( (\ga |V_C |\gb - \bm{1}_i]^{(m)} - (\ga \, |V_C |\gb - \bm{1}_i]^{(m+1)} \Bigr) .
\end{align}
Contrast it to the corresponding relation for a 3-center 2-particle integral with (Mulliken) ket involving a primitive s-type Gaussian AO at $\mathbf{C}$, obtained from Eq. (39) in Ref. \citenum{VRG:obara:1986:JCP} by substitutions $\eta \to \zeta_c, \mathbf{Q}\to\mathbf{C}$ (see also Eq. (3) in Ref. \citenum{VRG:asadchev:2023:JCTC}):
\begin{align}
\label{eq:vrr_ab_0}
    (\ga + \bm{1}_i \, \gb |\gzeroC)^{(m)} = & (\mathbf{PA})_i (\ga \, \gb |\gzeroC)^{(m)} -
    \frac{\rho}{\zeta}(\mathbf{PC})_i (\ga \, \gb |\gzeroC)^{(m+1)} \nonumber \\
    & + \frac{a_i}{2\zeta} \Bigl( (\ga - \bm{1}_i \, \gb |\gzeroC)^{(m)} - \frac{\rho}{\zeta} (\ga - \bm{1}_i \, \gb |\gzeroC)^{(m+1)}  \Bigr) \nonumber \\
    & + \frac{b_i}{2\zeta} \Bigl( (\ga \, \gb - \bm{1}_i|\gzeroC)^{(m)} - \frac{\rho}{\zeta} (\ga \, \gb - \bm{1}_i |\gzeroC)^{(m+1)} \Bigr) ,
\end{align}
with $\rho \equiv \zeta \zeta_c / (\zeta + \zeta_c)$.
Let us introduce rescaled auxiliary 3-center integrals:
\begin{align}
\label{eq:bar_ab_0}
\overline{(\ga \, \gb |\gzeroC)}^{(m)} \equiv 
\left(\frac{\rho}{\zeta}\right)^m (\ga \, \gb |\gzeroC)^{(m)};
\end{align}
note that the rescaling does not affect the target integrals ($m=0$):
\begin{align}
  \overline{(\ga \, \gb |\gzeroC)}^{(0)} = (\ga \, \gb |\gzeroC)^{(0)}.
\end{align}
Rewritting \cref{eq:vrr_ab_0} in terms of the rescaled integrals makes it isomorphic to \cref{eq:vrr_ab_V}:
\begin{align}
\label{eq:vrr_bar_ab_0}
\overline{(\ga + \bm{1}_i \, \gb |\gzeroC)}^{(m)} = & (\mathbf{PA})_i \overline{(\ga \, \gb |\gzeroC)}^{(m)} -
    (\mathbf{PC})_i \overline{(\ga \, \gb |\gzeroC)}^{(m+1)} \nonumber \\
    & + \frac{a_i}{2\zeta} \Bigl( \overline{(\ga - \bm{1}_i \, \gb |\gzeroC)}^{(m)} - \overline{(\ga - \bm{1}_i \, \gb |\gzeroC)}^{(m+1)}  \Bigr) \nonumber \\
    & + \frac{b_i}{2\zeta} \Bigl( \overline{(\ga \, \gb - \bm{1}_i|\gzeroC)}^{(m)} - \overline{(\ga \, \gb - \bm{1}_i |\gzeroC)}^{(m+1)} \Bigr) .
\end{align}
Therefore a superposition of $(\ga | V_C | \gb )$ with an arbitrary number of $(\ga \, \gb |\gzeroC)$ can be computed using Eq. \cref{eq:vrr_ab_V} by redefining the starting integrals over $l=0$ Gaussians appropriately.
Namely (see Eqs. (A20) and (44) in Ref. \citenum{VRG:obara:1986:JCP}, respectively):
\begin{align}
\label{eq:V0_OSRR}
(\gzeroA  | V_C | \gzeroB)^{(m)} = & - 2 Z_C \left(\frac{\zeta}{\pi}\right)^{1/2} (\gzeroA||\gzeroB) F_m(T) \\
\label{eq:3c0_OSRR}
\overline{(\gzeroA \, \gzeroB |\gzeroC)}^{(m)} = & 
2 \left(\frac{\zeta}{\pi}\right)^{1/2}
(\gzeroA||\gzeroB) F_m\left(T \frac{\zeta_c}{\zeta + \zeta_c} \right) \left(\frac{\zeta_c}{\zeta + \zeta_c} \right)^{m + \frac{1}{2}} (\gzeroC||) .
\end{align}
$(\gzeroC||)$ is the integral of $\phi_{\mathbf{0}_C}$ over the entire space, hence $(\gzeroC||)=1$ for unit-charge primitives used to represent the electronic contribution of SAP.
Combining \cref{eq:V0_OSRR} and \cref{eq:3c0_OSRR} according to \cref{eq:v-sap-r2,eq:thetaC} yields the SAP OS starting integrals:
\begin{align}
\label{eq:V0_SAP_OSRR}
(\gzeroA  | V^\text{SAP}_C | \gzeroB)^{(m)} = & 2 \left(\frac{\zeta}{\pi}\right)^{1/2}
(\gzeroA||\gzeroB) \left[ - Z_C  F_m \left(T \right) + \sum_k c_k F_m\left(T \frac{\alpha_k}{\zeta + \alpha_k} \right) \left(\frac{\alpha_k}{\zeta + \alpha_k} \right)^{m + \frac{1}{2}} \right] \nonumber \\
= & - 2 Z_C \left(\frac{\zeta}{\pi}\right)^{1/2}
(\gzeroA||\gzeroB) \left[ F_m \left(T \right) - \sum_k \tilde{c}_k F_m\left(T \frac{\alpha_k}{\zeta + \alpha_k} \right) \left(\frac{\alpha_k}{\zeta + \alpha_k} \right)^{m + \frac{1}{2}} \right].
\end{align}
Comparison of \cref{eq:V0_OSRR} to \cref{eq:V0_SAP_OSRR} confirms \cref{eq:BoysReplacementSAP}.

\cref{eq:BoysReplacementSAP} can be used in any evaluation method that follows the Boys route. In the McMurchie-Davidson (MD) method, it is possible to go further. Since evaluation of nuclear attraction integrals in the MD framework makes it possible to perform summation over nuclei inside the innermost loops,\cite{VRG:asadchev:2025:JPCA} it is therefore possible to combine nuclear and electronic contributions to SAP from all centers in the starting integrals. See the discussion of the nuclear attraction integral evaluation in Ref. \citenum{VRG:asadchev:2025:JPCA} for further details.

Our method for evaluating SAP integrals was implemented in the open-source Gaussian AO integral engines Libint2\cite{Libint2} and LibintX\cite{VRG:asadchev:2024:JCP,VRG:asadchev:2025:JPCA}. The described method for evaluating the SAP Gaussian integrals is not only an esthetically pleasing choice.
By evaluating nuclear and electronic contributions to SAP together, it can reduce the numerical roundoff error in the evaluation of SAP from nuclear (1-electron) and electronic (2-electron) contributions separately and avoid the divergence of the nuclear and electronic components of the potential in thermodynamic limit. This approach can be trivially extended to the case of Gaussian models of finite nuclei in relativistic computations\cite{VRG:visscher:1997:ADaNDT} 
by replacing the nuclear point charge contribution ($F_m(T)$) on the right-hand side of \cref{eq:BoysReplacementSAP} with the corresponding contribution $\left( \frac{\xi}{\zeta + \xi} \right)^{m+\frac{1}{2}} F_m\left(T \frac{\xi}{\zeta + \xi} \right)$ referring to a Gaussian nuclear density with exponent $\xi$ (see Eqs. (11-13) in Ref. \citenum{VRG:visscher:1997:ADaNDT}).
Lastly, the single-shot evaluation of SAP simplifies the evaluation of its various derivatives, as needed, for example, in the recently proposed SAP-based eXact 2-Component (X2C) method\cite{surjuse2026sapx2coptimallysimpletwocomponentrelativistic}; the additional SAP-including relativistic 1-electron integrals can be obtained from those needed for 1eX2C via the \cref{eq:BoysReplacementSAP} replacement.

This work was supported by the U.S. Department of Energy via award DE-SC0022327. The development of the \code{Libint}  and \code{LibintX} software libraries is supported by the Office of Advanced Cyberinfrastructure, National Science Foundation (Award OAC-2103738).

\bibliography{vrgrefs, misc}

\end{document}